\documentclass[12pt]{article}
\usepackage{graphicx}
\def\hybrid{\topmargin 0pt      \oddsidemargin 0pt
        \headheight 0pt \headsep 0pt
       \voffset-1cm
        \textwidth 6.25in       % A4 paper
       \textheight 9.5in       % A4 paper
        \marginparwidth 0.0in
        \parskip 5pt plus 1pt   \jot = 1.5ex}
\catcode`\@=11
\def\marginnote#1{}

\newcount\hour
\newcount\minute
\newtoks\amorpm
\hour=\time\divide\hour by60
\minute=\time{\multiply\hour by60 \global\advance\minute by-\hour}
\edef\standardtime{{\ifnum\hour<12 \global\amorpm={am}%
        \else\global\amorpm={pm}\advance\hour by-12 \fi
        \ifnum\hour=0 \hour=12 \fi
        \number\hour:\ifnum\minute<10 0\fi\number\minute\the\amorpm}}
\edef\militarytime{\number\hour:\ifnum\minute<10 0\fi\number\minute}

\def\draftlabel#1{{\@bsphack\if@filesw {\let\thepage\relax
   \xdef\@gtempa{\write\@auxout{\string
      \newlabel{#1}{{\@currentlabel}{\thepage}}}}}\@gtempa
   \if@nobreak \ifvmode\nobreak\fi\fi\fi\@esphack}
        \gdef\@eqnlabel{#1}}
\def\@eqnlabel{}
\def\@vacuum{}
\def\draftmarginnote#1{\marginpar{\raggedright\scriptsize\tt#1}}

\def\draftlabel#1{{\@bsphack\if@filesw {\let\thepage\relax
   \xdef\@gtempa{\write\@auxout{\string
      \newlabel{#1}{{\@currentlabel}{\thepage}}}}}\@gtempa
   \if@nobreak \ifvmode\nobreak\fi\fi\fi\@esphack}
        \gdef\@eqnlabel{#1}}
\def\@eqnlabel{}
\def\@vacuum{}
\def\draftmarginnote#1{\marginpar{\raggedright\scriptsize\tt#1}}

\def\draft{\oddsidemargin -.5truein
        \def\@oddfoot{\sl preliminary draft \hfil
        \rm\thepage\hfil\sl\today\quad\militarytime}
        \let\@evenfoot\@oddfoot \overfullrule 3pt
        \let\label=\draftlabel
        \let\marginnote=\draftmarginnote
   \def\@eqnnum{(\theequation)\rlap{\kern\marginparsep\tt\@eqnlabel}%
\global\let\@eqnlabel\@vacuum}  }

\def\numberbysection{\@addtoreset{equation}{section}
        \def\theequation{\thesection.\arabic{equation}}}

\def\underline#1{\relax\ifmmode\@@underline#1\else
        $\@@underline{\hbox{#1}}$\relax\fi}

\def\titlepage{\@restonecolfalse\if@twocolumn\@restonecoltrue\onecolumn
     \else \newpage \fi \thispagestyle{empty}\c@page\z@
        \def\thefootnote{\fnsymbol{footnote}} }

\def\endtitlepage{\if@restonecol\twocolumn \else  \fi
        \def\thefootnote{\arabic{footnote}}
        \setcounter{footnote}{0}}  %\c@footnote\z@ }
\relax

\numberbysection
\hybrid

\newfont{\Bbb}{msbm10 scaled 1\@ptsize00}
\newfont{\Bbbb}{msbm7 scaled 1\@ptsize00}

\newcommand{\DDD}{\raise-1pt\hbox{$\mbox{\Bbbb D}$}}

\newcommand{\UUU}{\raise-1pt\hbox{$\mbox{\Bbbb U}$}}

\newcommand{\ZZ}{\mbox{\Bbb Z}}
\newcommand{\z}{\raise-1pt\hbox{$\mbox{\Bbbb Z}$}}

\newcommand{\SSS}{\mbox{\Bbb S}}
\newcommand{\sss}{\raise-1pt\hbox{$\mbox{\Bbbb S}$}}

\def\beq{\begin{equation}}
\def\eeq{\end{equation}}
\def\p{\partial}

\newtheorem{theorem}{Theorem}[section]

\newtheorem{lemma-definition}{Lemma-Definition}[section]

\newtheorem{definition}{Definition}[section]
\newtheorem{proposition}{Proposition}[section]

\def\square{\hfill
{\vrule height6pt width6pt depth1pt} \break \vspace{.01cm}}

\begin{document}

\begin{titlepage}

\title{Constrained Toda hierarchy and turning points of the Ruijsenaars-Schneider model}

\author{I. Krichever\thanks{Skolkovo Institute of Science and
Technology, 143026, Moscow, Russia and
National Research University Higher School of
Economics,
20 Myasnitskaya Ulitsa, Moscow 101000, Russia, and
Columbia University, New York, USA;
e-mail: krichev@math.columbia.edu}
\and
A.~Zabrodin\thanks{
Skolkovo Institute of Science and Technology, 143026, Moscow, Russia and
National Research University Higher School of Economics,
20 Myasnitskaya Ulitsa,
Moscow 101000, Russia and
ITEP NRC KI, 25
B.Cheremushkinskaya, Moscow 117218, Russia;
e-mail: zabrodin@itep.ru}}

\date{September 2021}
\maketitle

\vspace{-7cm} \centerline{ \hfill ITEP-TH-28/21}\vspace{7cm}

\begin{abstract}

We introduce a new integrable hierarchy of nonlinear differential-difference equations which
we call constrained Toda hierarchy (C-Toda). It can be regarded as a certain subhierarchy
of the 2D Toda lattice obtained by imposing the constraint $\bar {\cal L}={\cal L}^{\dag}$
on the two Lax operators (in the symmetric gauge). We prove the existence of the tau-function
of the C-Toda hierarchy and show that it is the square root of the 2D Toda lattice tau-function.
In this and some other respects the C-Toda is a Toda analogue of the CKP hierarchy.
It is also shown that zeros of the tau-function of elliptic solutions satisfy the dynamical
equations of the Ruijsenaars-Schneider model restricted to turning points in the phase
space. The spectral curve has holomorphic involution which interchange the marked points
in which the Baker-Akhiezer function has essential singularities.

\end{abstract}

\end{titlepage}

\vspace{5mm}

%

%\newpage
\tableofcontents

\vspace{5mm}

\section{Introduction}

The 2D Toda lattice hierarchy \cite{UT84}
is perhaps the most fundamental in the theory of integrable
systems. The commuting flows of the hierarchy are parametrized by infinite sets of time
variables ${\bf t}=\{t_1, t_2, t_3, \ldots \}$ (``positive times'') and $\bar {\bf t}=\{\bar t_1,
\bar t_2, \bar t_3, \ldots \}$ (``negative times''), together with the ``zeroth time'' $t_0=x$.
Equations of the hierarchy are differential in the times ${\bf t}$, $\bar {\bf t}$ and
difference in $x$ with a lattice spacing $\eta$.
A common solution is provided by the tau-function $\tau = \tau (x, {\bf t}, \bar {\bf t})$
which satisfies an infinite set of bilinear differential-difference
equations of Hirota type \cite{DJKM83,JM83}. All dependent variables are expressed through
the tau-function in one or another way.

Equally fundamental is the Kadomtsev-Petviashvili (KP) hierarchy
with independent variables  ${\bf t}=\{t_1, t_2, t_3, \ldots \}$ which can be regarded
as a subhierarchy of the 2D Toda lattice obtained by fixing the times $\bar {\bf t}$ and $t_0$.
Equations of the KP hierarchy are purely differential.

Many if not all known integrable nonlinear partial differential and difference equations
are reductions or special cases of the 2D Toda lattice and KP hierarchies. Remarkably, they
also contain most of the known finite-dimensional many-body integrable systems. For example,
solutions of the KP hierarchy which are elliptic functions of $t_1$ with $N$ poles
in the fundamental domain (zeros of the tau-function) give rise to the $N$-body
elliptic Calogero-Moser system \cite{Calogero71,Moser75,OP81}:
zeros of the tau-function as functions of $t_2$
move as Calogero-Moser particles (see \cite{AMM77,CC77,Krichever78,Krichever80}
and \cite{Z19} for a review). Later it was shown that this correspondence can be extended to
all commuting flows of the hierarchy: the $t_j$-dynamics of zeros of the tau-function
is the same as the Calogero-Moser dynamics with respect to the higher Hamiltonian
$H_j$ (see \cite{Shiota94,Z19a,PZ21a}). In their turn, poles of solutions of the 2D Toda lattice
hierarchy which are elliptic functions of $t_0$ move as particles of the Ruijsenaars-Schneider
model \cite{RS86,Ruij87} which can be regarded
as a relativistic extension of the Calogero-Moser model (see \cite{KZ95,PZ21}).

Given an integrable hierarchy with a space of solutions
${\cal M}$, one can define a subhierarchy by imposing some constraints
which restrict the space of solutions to ${\cal X}\subset {\cal M}$. In known examples
the constraints are preserved by only a part of the commuting flows of the hierarchy and
are destroyed by the other part, so these time variables should be frozen.

Well known examples of such situation are provided by the B- and C-versions of the KP hierarchy
(BKP and CKP). In particular, the CKP hierarchy is introduced by imposing the constraint
${\cal L}^{\dag} =-{\cal L}$ on the Lax operator of the KP hierarchy, where the operation
${}^{\dag}$ is defined as $(f(x)\circ \p_x^n)^{\dag}=(-\p_x)^n \circ f(x)$. The constraint
is preserved by the ``odd'' flows and is destroyed by the ``even'' ones, so one should
fix ``even'' times to zero values: $t_{2j}=0$ for all $j$.
The CKP hierarchy was introduced in the paper
\cite{DJKM81} and later different aspects of it were discussed in \cite{DM-H09,CW13,CH14,LOS12}.
Recently, in \cite{KZ21}, a characterization of the CKP hierarchy in terms of KP
tau-function was obtained: it was shown that the KP tau-functions that provide solutions
of the CKP hierarchy (with frozen ``even'' times) are characterized by the condition
\beq\label{ckp1}
\p_{t_2}\log \tau \Bigr |_{t_{2j}=0}=0.
\eeq
This condition makes sense as defining ``turning points'' for zeros $x_i$ of the tau-function
in the variable $x=t_1$: $\p_{t_2}x_i=0$ (the velocities vanish). For elliptic solutions,
the zeros of the tau-function move as particles of the elliptic Calogero-Moser system, so
the condition (\ref{ckp1}) indeed defines the submanifold of turning points in the phase space,
where all momenta $p_i=2\p_{t_2}x_i$ are equal to zero. General algebraic-geometrical
solutions to the CKP hierarchy are obtained starting from algebraic curves which have
a holomorphic involution, with the marked point on the curve (the point where the Baker-Akhiezer
function has essential singularity) being a fixed point of the involution.

Moreover, one can prove that the CKP hierarchy possesses its own tau-function
$\tau^{\rm CKP}$ which is a function of the ``odd'' times only, and this tau-function
is given by square root of the KP tau-function restricted to the turning points.

In this paper, we suggest a Toda analogue of this story. To wit,
we introduce a subhierarchy of the 2D Toda lattice which is related to it in the way
much similar to the relation between the CKP and KP hierarchies
We call it C-Toda hierarchy\footnote{It is different from what is 
called Toda hierarchy of C-type in \cite{UT84}.}
(``C'' is from ``constrained'' and simultaneously points to the similarity with CKP.).
The constraint connects the two
pseudo-difference Lax operators ${\cal L}$, $\bar {\cal L}$ as follows:
\beq\label{ctoda1}
\bar {\cal L}={\cal L}^{\dag}
\eeq
(in the symmetric gauge). This constraint is preserved by the flows $\p_{t_j}-\p_{\bar t_j}$
and is destroyed by the flows $\p_{t_j}+\p_{\bar t_j}$, so one should fix
$t_j+\bar t_j=0$ and vary only the times $T_j=\frac{1}{2}(t_j-\bar t_j)$. We show that
solutions to the C-Toda hierarchy among all solutions to the 2D Toda lattice are
characterized by the condition
\beq\label{ctoda2}
(\p_{t_1}+\p_{\bar t_1})\log \tau \Bigr |_{t_j+\bar t_j=0}=0.
\eeq
Similarly to the CKP case, this condition makes sense
as defining ``turning points'' for zeros $x_i$ of the tau-function in the variable $x$
(the ``zeroth time'' of the 2D Toda lattice): $(\p_{t_1}+\p_{\bar t_1})x_i=0$.
For elliptic solutions,
the zeros of the tau-function move as particles of the elliptic Ruijsenaars-Schneider
system, so
the condition (\ref{ctoda2}) indeed defines the submanifold of turning points in the phase space.

We also prove that the C-Toda hierarchy possesses its own tau-function
$\tau^{C}$ which is a function of the times $T_j$ only, and this tau-function
is given by square root of the 2D Toda lattice  tau-function restricted to the turning points.

The analogies between the CKP and C-Toda hierarchies are summarized in the table:

\begin{center}

\begin{tabular}{l|c|c}
& CKP & C-Toda \\ &&\\
\hline
&& \\
Evolution times & $\begin{array}{c}t_1, t_3, t_5, \ldots ;\\ t_{2j}=0\end{array}$ &
$\begin{array}{c}t_1\! -\! \bar t_1, t_2\! -\! \bar t_2, t_3\! -\! \bar t_3, \ldots ;\\
t_j\! +\! \bar t_j=0 \end{array}$\\
&& \\
\hline
&& \\
$\begin{array}{l}\mbox{Constraints}\\ \mbox{for $L$-operators}\end{array}$
& ${\cal L}^{\dag}=-{\cal L}$ &
$\bar {\cal L}={\cal L}^{\dag}$\\
&&\\
\hline
&&\\
Tau-functions & $\tau^{\rm CKP}=\sqrt{\vphantom{I^{I^I}}\tau^{\rm KP}}$ &
$\tau^{\rm C-Toda}=\sqrt{\vphantom{I^{I^I}}\tau^{\rm Toda}}$\\
&&\\
\hline
&&\\
$\begin{array}{l}\mbox{Turning points}\\ \mbox{conditions}\end{array}$ &
$\displaystyle{\p_{t_2}\log \tau^{\rm KP}\Bigr |_{t_{2j}=0}=0}$ &
$\displaystyle{(\p_{t_1}+\p_{\bar t_1})\log \tau^{\rm Toda}\Bigr |_{t_j+\bar t_j=0}=0}$\\
&&\\
\hline
&&\\
Bilinear relations & $\displaystyle{
\oint_{C_{\infty}}\! \!\!\! \psi ({\bf t}, k)\psi ({\bf t}', -k)dk\! =\! 0}$ &
$\displaystyle{\Bigl (\oint_{C_{\infty}}\! \!\! -\!\oint_{C_0}\Bigr )
\psi ({\bf t}, k)\psi ({\bf t}', k^{-1})\frac{dk}{k}\! =\! 0}$\\
&&\\
\hline
&&\\
Algebraic curves & $\begin{array}{c}\mbox{involution $\iota$,}
\\ \iota P_{\infty}=P_{\infty}\end{array}$ & $\begin{array}{c}\mbox{involution $\iota$,}
\\ \iota P_{\infty}=P_0, \iota P_0=P_{\infty}\end{array}$\\
&&\\
\end{tabular}

\end{center}

The paper is organized as follows. In section 2.1 we briefly review the 2D Toda lattice
hierarchy. In section 2.2 the constrained Toda hierarchy (C-Toda) is introduced and in section
2.3 we prove the existence of the tau-function for this hierarchy. Section 3 is devoted to the
elliptic Ruijsenaars-Schneider model. We show that elliptic solutions of the C-Toda
hierarchy generate the Ruijsenaars-Schneider dynamics of their poles (zeros of the
tau-function) restricted to the subspace in the phase space corresponding to turning points.
We also prove that the spectral curve of the Lax matrix of the Ruijsenaars-Schneider model
for turning points admits a holomorphic involution.

\section{Constrained Toda hierarchy}

\subsection{2D Toda lattice}

First of all, we briefly review the 2D Toda lattice hierarchy following \cite{UT84}.
Let us consider the pseudo-difference Lax operators
\beq\label{mkp1}
{\cal L}=e^{\eta \p_x}+\sum_{k\geq 0}U_k(x) e^{-k\eta \p_x}, \quad
\bar {\cal L}=c(x)e^{-\eta \p_x}+\sum_{k\geq 0}\bar U_k(x) e^{k\eta \p_x},
\eeq
where $e^{\eta \p_x}$ is the shift operator acting as
$e^{\pm \eta \p_x}f(x)=f(x\pm \eta )$ and
the coefficient functions $U_k$, $\bar U_k$
are functions of $x$, ${\bf t}$, $\bar {\bf t}$.
The Lax equations are
\beq\label{mkp2}
\p_{t_m}{\cal L}=[{\cal B}_m, {\cal L}], \quad
\p_{t_m}\bar {\cal L}=[{\cal B}_m, \bar {\cal L}]
\qquad {\cal B}_m=({\cal L}^m)_{\geq 0},
\eeq
\beq\label{mkp2a}
\p_{\bar t_m}{\cal L}=[\bar {\cal B}_m, {\cal L}], \quad
\p_{\bar t_m}\bar {\cal L}=[\bar {\cal B}_m, \bar {\cal L}]
\qquad \bar {\cal B}_m=(\bar {\cal L}^m)_{< 0}.
\eeq
Here and below, given a subset $\SSS \subset \ZZ$, we denote
$\displaystyle{\Bigl (\sum_{k\in \z} U_k e^{k\eta \p_x}\Bigr )_{\sss}=
\sum_{k\in \sss} U_k e^{k\eta \p_x}}$.
For example, ${\cal B}_1=e^{\eta \p_x}+U_0(x)$,
$\bar {\cal B}_1=c(x)e^{-\eta \p_x}$.
An equivalent formulation is through the zero
curvature (Zakharov-Shabat) equations
\beq\label{mkp3}
\p_{t_n}{\cal B}_m -\p_{t_m}{\cal B}_n +[{\cal B}_m, {\cal B}_n]=0,
\eeq
\beq\label{mkp3a}
\p_{\bar t_n}{\cal B}_m -\p_{t_m}\bar {\cal B}_n
+[{\cal B}_m, \bar {\cal B}_n]=0,
\eeq
\beq\label{mkp3b}
\p_{\bar t_n}\bar {\cal B}_m -\p_{\bar t_m}\bar {\cal B}_n
+[\bar {\cal B}_m, \bar {\cal B}_n]=0.
\eeq

For example, putting
\beq\label{toda1}
c(x)=e^{\varphi (x)-\varphi (x-\eta )},
\eeq
we have
from (\ref{mkp3a}) at $m=n=1$:
\beq\label{toda2}
\p_{t_1}\p_{\bar t_1}\varphi (x)=e^{\varphi (x)-\varphi (x-\eta )}-
e^{\varphi (x+\eta )-\varphi (x)}.
\eeq
This is the famous 2D Toda lattice equation.

Note that from (\ref{mkp2}), (\ref{mkp2a}) it follows that
\beq\label{toda5}
\p_{t_m}\varphi =({\cal L}^m)_0, \quad
\p_{\bar t_m}\varphi =-(\bar {\cal L}^m)_0.
\eeq

The zero curvature equations
are compatibility conditions for the auxiliary linear problems
\beq\label{mkp6}
\p_{t_m}\psi ={\cal B}_m (x)\psi , \quad
\p_{\bar t_m}\psi =\bar {\cal B}_m (x)\psi ,
\eeq
where the wave function $\psi$ depends on a spectral parameter $k$:
$\psi =\psi (x,{\bf t}, \bar {\bf t}; k)$. The wave function has the
following expansion in powers of
$k$:
\beq\label{toda3}
\psi (x,{\bf t}, \bar {\bf t}; k)=\left \{
\begin{array}{l}
\displaystyle{ k^{x/\eta}e^{\xi ({\bf t}, k)}\Bigl (1+\sum_{s\geq 1} \xi_s (x)k^{-s}\Bigr )},
\quad k\to \infty ,
\\ \\
\displaystyle{
k^{x/\eta}e^{\xi (\bar {\bf t}, k^{-1})+\varphi (x)}
\Bigl (1+\sum_{s\geq 1} \chi_s (x)k^{s}\Bigr )},
\quad k\to 0,
\end{array}
\right.
\eeq
where
\beq\label{mkp8}
\xi ({\bf t}, k)=\sum_{j\geq 1}t_j k^j.
\eeq
The wave function satisfies the linear equation
\beq\label{inv3}
\p_{t_1} \psi (x,k)=\psi (x+\eta ,k)+v(x)\psi (x, k),
\eeq
where
$v(x)=U_0(x)$.

The wave operators are pseudo-difference operators of the form
\beq\label{mkp9}
\begin{array}{l}
{\cal W}(x)=1+\xi_1(x)e^{-\eta \p_x}+\xi_2(x)e^{-2\eta \p_x}+\ldots \,
\\ \\
\bar {\cal W}(x)=e^{\varphi (x)}(1+\chi_1(x)e^{-\eta \p_x}+\chi_2(x)e^{-2\eta \p_x}+\ldots )
\end{array}
\eeq
with the same coefficient functions $\xi_j$, $\chi_j$ as in
(\ref{toda3}), then the wave function can be written as
\beq\label{mkp10}
\begin{array}{l}
\psi = {\cal W}(x)k^{x/\eta} e^{\xi ({\bf t}, k)}, \quad k\to \infty ,
\\ \\
\psi = \bar {\cal W}(x)k^{x/\eta} e^{\xi (\bar {\bf t}, k^{-1})}, \quad k\to 0.
\end{array}
\eeq

The dual wave function $\psi^{*}$ is defined by
\beq\label{mkp11}
\psi^{*}=({\cal W}^{\dag}(x))^{-1}k^{-x/\eta} e^{-\xi ({\bf t}, k)}, \quad k\to \infty ,
\eeq
where the adjoint difference operator is defined according to the rule
$(f(x) \circ e^{n\eta \p_x})^{\dag}=e^{-n\eta \p_x}\circ f(x)$. The auxiliary linear
problems for the dual wave function have the form
\beq\label{mkp12}
-\p_{t_m}\psi ^{*}={\cal B}_{m}^{\dag}(x)\psi^{*}.
\eeq

The Lax operators (\ref{mkp1}) are obtained by ``dressing'' of the shift operators
by ${\cal W}$, $\bar {\cal W}$:
\beq\label{toda4}
{\cal L}={\cal W}e^{\eta \p_x}{\cal W}^{-1}, \quad
\bar {\cal L}=\bar {\cal W}e^{-\eta \p_x}\bar {\cal W}^{-1}.
\eeq

So far we have used the standard gauge in which
the coefficient of the first term of ${\cal L}$ is fixed to be $1$. In fact
there is a family of gauge transformations with $g=e^{\alpha \varphi (x)}$
\cite{Takebe1,Takebe2}:
$$
{\cal L}\to g^{-1}{\cal L}g , \quad \bar {\cal L}\to g^{-1}\bar {\cal L}g,
$$
$$
{\cal B}_n \to g^{-1}{\cal B}_n g -g^{-1}\p_{t_n}g, \quad
\bar {\cal B}_n \to g^{-1}\bar {\cal B}_n g -g^{-1}\p_{\bar t_n}g
$$
of which $\alpha =0$ corresponds to the standard gauge ${\cal L}={\cal L}^{(0)}$,
$\bar {\cal L}=\bar {\cal L}^{(0)}$. At $\alpha =\frac{1}{2}$ we have the so-called
symmetric gauge:
\beq\label{toda6}
{\cal L}^s=c^s(x)e^{\eta \p_x}+\sum_{k\geq 0}U^s_k(x) e^{-k\eta \p_x}, \quad
\bar {\cal L}^s=c^s(x\! -\! \eta )e^{-\eta \p_x}+\sum_{k\geq 0}\bar U^s_k(x) e^{k\eta \p_x},
\eeq
\beq\label{toda7}
c^s(x)=e^{\frac{1}{2}(\varphi (x+\eta )-\varphi (x))}.
\eeq
Hereafter, we write simply ${\cal L}^s$, $\bar {\cal L}^s$ instead of
${\cal L}^{(1/2)}$, $\bar {\cal L}^{(1/2)}$ for brevity. In the symmetric gauge, the
generators of the $t_m$- and $\bar t_m$-flows ${\cal B}_m$, $\bar {\cal B}_m$ are
\beq\label{toda8}
{\cal B}^s_m=(({\cal L}^s)^m)_{>0}+\frac{1}{2}\, (({\cal L}^s)^m)_{0} , \quad
\bar {\cal B}^s_m=((\bar {\cal L}^s)^m)_{<0}+\frac{1}{2}\, ((\bar {\cal L}^s)^m)_{0}.
\eeq
Similarly to (\ref{toda4}), the Lax operators ${\cal L}^s$, $\bar {\cal L}^s$
are obtained by dressing of the shift operators:
\beq\label{toda9}
{\cal L}^s={\cal W}^s e^{\eta \p_x}({\cal W}^s)^{-1}, \quad
\bar {\cal L}^s=\bar {\cal W}^s e^{-\eta \p_x}(\bar {\cal W}^s)^{-1},
\eeq
where the wave operators are
\beq\label{toda10}
{\cal W}^s(x)=e^{-\frac{1}{2}\, \varphi (x)}{\cal W}, \quad
\bar {\cal W}^s(x)=e^{-\frac{1}{2}\, \varphi (x)}\bar {\cal W}.
\eeq
We also note that the wave functions are given by
\beq\label{toda11}
\psi (x, k)=e^{\frac{1}{2}\, \varphi (x)} \bar {\cal W}^s
(x)k^{x/\eta}e^{\xi (\bar {\bf t}, k^{-1})},
\quad k\to 0,
\eeq
\beq\label{toda12}
\psi^* (x, k)=e^{-\frac{1}{2}\, \varphi (x)}
({\cal W}^{s \dag}(x))^{-1}k^{-x/\eta}e^{-\xi ({\bf t}, k)},
\quad k\to \infty .
\eeq

A common solution to the 2D Toda lattice hierarchy is provided by the tau-function
$\tau =\tau (x, {\bf t}, \bar {\bf t})$ \cite{DJKM83,JM83}.
The tau-function satisfies the bilinear relation
\beq\label{bil}
\begin{array}{c}
\displaystyle{\oint_{C_{\infty}}k^{\frac{x-x'}{\eta}-1}e^{\xi ({\bf t}, k)-\xi ({\bf t}', k)}
\tau \Bigl (x, {\bf t}-[k^{-1}], \bar {\bf t}\Bigr )
\tau \Bigl (x'+\eta , {\bf t}'+[k^{-1}], \bar {\bf t}'\Bigr )dk}
\\ \\
\displaystyle{=\, \oint_{C_{0}}k^{\frac{x-x'}{\eta}-1}
e^{\xi (\bar {\bf t}, k^{-1})-\xi (\bar {\bf t}', k^{-1})}
\tau \Bigl (x+\eta , {\bf t}, \bar {\bf t}-[k]\Bigr )
\tau \Bigl (x' , {\bf t}', \bar {\bf t}'+[k]\Bigr )dk
}
\end{array}
\eeq
valid for all $x, x'$, ${\bf t}, {\bf t}'$, $\bar {\bf t}, \bar {\bf t}'$.
It is assumed that $x-x'\in \eta \ZZ$.
The integration contour $C_{\infty}$ in the left hand side
is a big circle around infinity separating the singularities
coming from the exponential factor from those coming from the tau-functions.
The integration contour $C_{0}$ in the right hand side
is a small circle around zero separating the singularities
coming from the exponential factor from those coming from the tau-functions.
The bilinear relation (\ref{bil}) encodes all differential-difference equations of the
hierarchy.

Setting $x-x'=\eta$, $t_n -t'_n=\frac{1}{n}a^{-n}$, $\bar t_n -\bar t'_n=\frac{1}{n}b^{-n}$
in (\ref{bil}) and taking the residues, we get the 3-term bilinear equation of the
Hirota-Miwa type:
\beq\label{bil2}
\begin{array}{l}
\tau (x, {\bf t}-[a^{-1}], \bar {\bf t})\tau (x, {\bf t}, \bar {\bf t}-[b^{-1}])
-\tau (x, {\bf t}, \bar {\bf t})\tau (x, {\bf t}-[a^{-1}], \bar {\bf t}-[b^{-1}])
\\ \\
\phantom{aaaaaaaaaaa}
=(ab)^{-1}
\tau (x-\eta, {\bf t}-[a^{-1}], \bar {\bf t})\tau (x+\eta, {\bf t}, \bar {\bf t}-[b^{-1}]).
\end{array}
\eeq

The functions $\varphi (x), U_0(x)$ are expressed through the tau-function as follows:
\beq\label{inv2}
\varphi (x)=\log \frac{\tau (x+\eta)}{\tau (x)},
\eeq
\beq\label{inv2aa}
U_0(x)=\p_{t_1} \log \frac{\tau (x+\eta)}{\tau (x)}=\p_{t_1}\varphi (x).
\eeq
The wave function $\psi (x, k)$ and its dual $\psi^*(x,k)$ are expressed
through the tau-function as follows \cite{UT84,DJKM83,JM83}:
\beq\label{inv15}
\begin{array}{l}
\displaystyle{
\psi (x, k)=k^{x/\eta}\exp (\sum_{j\geq 1} t_jk^j}\Bigr )
\, \frac{\tau \Bigl (x, {\bf t}-[k^{-1}], \bar {\bf t} \Bigr )}{\tau 
(x, {\bf t})},
\quad k\to \infty ,
\\ \\
\displaystyle{
\psi (x, k)=k^{x/\eta}\exp (\sum_{j\geq 1} \bar t_jk^{-j}}\Bigr )
\, \frac{\tau \Bigl (x+\eta , {\bf t}, \bar {\bf t}-[k] \Bigr )}{\tau 
(x, {\bf t})},
\quad k\to 0 ,
\\ \\
\displaystyle{
\psi^* (x, k)=k^{-x/\eta}\exp (-\sum_{j\geq 1} t_jk^j}\Bigr )
\, \frac{\tau \Bigl (x+\eta, {\bf t}+[k^{-1}],
\bar {\bf t} \Bigr )}{\tau (x+\eta, {\bf t})},
\quad k\to \infty ,
\\ \\
\displaystyle{
\psi^* (x, k)=k^{-x/\eta}\exp (-\sum_{j\geq 1} \bar t_jk^{-j}}\Bigr )
\, \frac{\tau \Bigl (x , {\bf t}, \bar {\bf t}+[k] \Bigr )}{\tau 
(x+\eta, {\bf t})},
\quad k\to 0 ,
\end{array}
\eeq
where
$$
{\bf t} \pm [k] =\Bigl \{ t_1 \pm k, t_2\pm \frac{1}{2}\, k^2, t_3\pm \frac{1}{3}\, k^3,
\ldots \Bigr \}.
$$
Taking into account formulas (\ref{inv15}), one can represent (\ref{bil}) as
a bilinear relation for the wave functions:
\beq\label{bil1}
\Bigl (\oint_{C_{\infty}}\! -\oint_{C_0}\Bigr )\,
\psi (x, {\bf t}, \bar {\bf t};k)\psi^* (x', {\bf t}', \bar {\bf t}';k)\frac{dk}{2\pi ik}=0,
\quad x-x'\in \eta \ZZ .
\eeq

\subsection{The C-Toda hierarchy}

The C-Toda hierarchy is defined by imposing the constraint
\beq\label{toda13}
\bar {\cal L}^s={\cal L}^{s \dag}
\eeq
(in the symmetric gauge). In the standard gauge, it looks as follows:
\beq\label{toda13a}
\bar {\cal L}e^{\varphi}=e^{\varphi} {\cal L}^{\dag}.
\eeq
This means that $\bar U^s_j(x)=U^s_j(x+j\eta )$ for $j\geq 0$.
In terms of the wave operators, this is equivalent to the constraint
\beq\label{toda14}
\bar {\cal W}^s{\cal W}^{s \dag}={\cal W}^s \bar {\cal W}^{s \dag}=1.
\eeq

It is important to note that not all time flows of the full Toda hierarchy are consistent
with the constraint. Let us introduce the following linear combinations of times:
\beq\label{toda15}
T_j = \frac{1}{2} (t_j-\bar t_j), \quad y_j = \frac{1}{2} (t_j+\bar t_j),
\eeq
then the corresponding vector fields are
\beq\label{toda16}
\p_{T_j}=\p_{t_j}-\p_{\bar t_j}, \quad
\p_{y_j}=\p_{t_j}+\p_{\bar t_j}.
\eeq
One can see that the $T_j$-flows preserve the constraint. Indeed, we have:
$$
\p_{t_j}(\bar {\cal L}^s -{\cal L}^{s\dag})=[{\cal B}_j^s, \bar {\cal L}^s]-
[{\cal B}_j^s, {\cal L}]^{s\dag}=[{\cal B}_j^s, \bar {\cal L}^s]+
[{\cal B}_j^{s\dag}, {\cal L}^{s\dag}]=
[{\cal B}_j^s +\bar {\cal B}_j^s, \bar {\cal L}^s]=(\p_{t_j}+\p_{\bar t_j})\bar {\cal L}^s.
$$
Similarly,
$$
\p_{\bar t_j}(\bar {\cal L}^s -{\cal L}^{s\dag})=
[\bar {\cal B}_j^s, \bar {\cal L}^s]-
[\bar {\cal B}_j^s, {\cal L}]^{s\dag}=[\bar {\cal B}_j^s, \bar {\cal L}^s]+
[\bar {\cal B}_j^{s\dag}, {\cal L}^{s\dag}]=
[{\cal B}_j^s +\bar {\cal B}_j^s, \bar {\cal L}^s]=(\p_{t_j}+\p_{\bar t_j})\bar {\cal L}^s,
$$
so
$$
(\p_{t_j}-\p_{\bar t_j})(\bar {\cal L}^s -{\cal L}^{s\dag})=
\p_{T_j}(\bar {\cal L}^s -{\cal L}^{s\dag})=0
$$
for all $T_j$. At the same time, the $y_j$-flows destroy the constraint, so we should
put $y_j=0$ for all $j$. The situation is similar to the embedding of the CKP hierarchy
into the KP one, where the constraint is preserved only by the ``odd'' times and all
``even'' times are fixed to be $0$.

Set
\beq\label{toda17}
{\cal A}_m ={\cal B}^s_m -\bar {\cal B}^s_m.
\eeq
In particular,
$$
{\cal A}_1=c^s(x)e^{\eta \p_x}-c^s(x-\eta )e^{-\eta \p_x},
$$
$$
{\cal A}_2=c^s(x)c^s(x+\eta )e^{2\eta \p_x}+c^s(x)(v(x)+v(x+\eta ))e^{\eta \p_x}
$$
$$
-c^s(x-\eta )(v(x)+v(x-\eta ))e^{-\eta \p_x}-
c^s(x-\eta )c^s(x-2\eta )e^{-2\eta \p_x},
$$
where $v(x)=U_0(x)=\frac{1}{2}\p_{T_1}\varphi (x)$.
The Zakharov-Shabat equations for the C-Toda hierarchy read
\beq\label{toda18}
[\p_{T_m}-{\cal A}_m, \, \p_{T_n}-{\cal A}_n]=0.
\eeq
The simplest equation is obtained at $m=1$, $n=2$. It reads:
\beq\label{toda19}
\begin{array}{l}
(\p_{T_2}-\p^2_{T_1})\varphi (x+\eta )-(\p_{T_2}+\p^2_{T_1})\varphi (x)
\\ \\
\phantom{aaaaa}
=2e^{\varphi (x)-\varphi (x-\eta )}-2e^{\varphi (x+2\eta )-\varphi (x+\eta )}+
\frac{1}{2} (\p_{T_1}\varphi (x+\eta ))^2 -\frac{1}{2} (\p_{T_1}\varphi (x ))^2.
\end{array}
\eeq

Equations (\ref{mkp10}) together with the constraints (\ref{toda14}) imply that
the dual wave function $\psi^*$ in the C-Toda hierarchy is expressed through the wave
function $\psi$ as follows:
\beq\label{toda20}
\psi^*(x, k)=e^{-\varphi (x)}\psi (x, k^{-1})\Bigr |_{t_j+\bar t_j=0}.
\eeq
The bilinear relation (\ref{bil1}) for the C-Toda hierarchy acquires the form
\beq\label{bil1a}
\Bigl (\oint_{C_{\infty}}\! -\oint_{C_0}\Bigr )\,
\psi (x, {\bf t}, \bar {\bf t};k)\psi (x', {\bf t}', \bar {\bf t}';k^{-1})\frac{dk}{2\pi ik}=0,
\quad x-x'\in \eta \ZZ ,
\eeq
where it is assumed that $t_j+\bar t_j=t'_j+\bar t'_j=0$.

Using relations (\ref{inv15}), we see that equation (\ref{toda20})
in terms of the tau-function reads
\beq\label{inv16}
\tau (x, {\bf t}, \bar {\bf t}-[k^{-1}])=\tau (x, {\bf t}+[k^{-1}], \bar {\bf t}) \quad
\mbox{at \,\,\, $t_k +\bar t_k=0$}.
\eeq
Expanding it in powers of $k$, we obtain, in the leading order:
\beq\label{inv17}
(\p_{t_1}+\p_{\bar t_1})\log \tau (x, {\bf t}, \bar {\bf t})=0
\quad
\mbox{at \,\,\, $t_k +\bar t_k=0$}.
\eeq
This is the necessary condition which should be obeyed by
the tau-function of the 2D Toda lattice in order to provide a solution to the C-Toda hierarchy.
We conjecture that this condition implies
\beq\label{inv17a}
(\p_{t_j}+\p_{\bar t_j})\log \tau (x, {\bf t}, \bar {\bf t})=0
\quad
\mbox{at \,\,\, $t_k +\bar t_k=0$}
\eeq
for all $j\geq 1$. In particular, we see that any solution of the 1D Toda 
hierarchy solves the constrained Toda hierarchy. 

\subsection{Tau-function of the C-Toda hierarchy}

The wave functions of the C-Toda hierarchy can be expressed through the tau-function
$\tau =\tau^T$ of the 2D Toda hierarchy according to formulas (\ref{inv15}). However,
one may ask whether there exists a tau-function $\tau^C$ of the C-Toda hierarchy which depends
on the time variables $T_j=\frac{1}{2}(t_j-\bar t_j)=t_j$ only
(hereafter, because at $t_j+\bar t_j=0$ we have $T_j=t_j$, we use the notation
$t_j$ for the time variables $T_j$). Below we show that the
answer is in the affirmative.

\begin{theorem}\label{theorem:tau}
There exists a function $\tau^C=\tau^C (x, {\bf t})$ such that
\beq\label{chi4b}
\psi (x, {\bf t}; k)=e^{\frac{1}{2}\varphi (x, {\bf t})}
\sqrt{\chi^2(x, {\bf t};k)-\chi^2(x\! -\! \eta , {\bf t};k)}, \quad k\to \infty ,
\eeq
\beq\label{chi4c}
\psi (x, {\bf t}; k^{-1})=e^{\frac{1}{2}\varphi (x, {\bf t})}
\sqrt{\bar \chi^2(x, {\bf t};k)-\bar \chi^2(x\! +\! \eta , {\bf t};k)}, \quad k\to \infty ,
\eeq
where
\beq\label{chi1a}
\chi (x, {\bf t};k)=k^{x/\eta}e^{\xi ({\bf t}, k) -\frac{1}{2}\varphi (x, {\bf t})}
\frac{\tau^C (x, {\bf t}-[k^{-1}])}{\tau^C (x, {\bf t})},
\eeq
\beq\label{chi2a}
\bar \chi (x, {\bf t};k)=k^{-x/\eta}e^{-\xi ({\bf t}, k)}
\frac{\tau^C (x+\eta , {\bf t}+[k^{-1}])}{\tau^C (x, {\bf t})},
\eeq
\beq\label{tau13a}
\varphi (x, {\bf t})=\log\left (\frac{\tau^C(x+\eta , {\bf t})}{\tau^C (x, {\bf t})}\right )^2.
\eeq
\end{theorem}

\begin{definition} The function $\tau^C=\tau^C (x, {\bf t})$ is called the tau-function of the
C-Toda hierarchy.
\end{definition}

\noindent
{\it Proof of Theorem \ref{theorem:tau}.}
The starting point of the proof
is the bilinear relation (\ref{bil1a}):
\beq\label{bil1b}
\Bigl (\oint_{C_{\infty}}\! -\oint_{C_0}\Bigr )\,
\psi (x, {\bf t}, -{\bf t};k)\psi (x', {\bf t}', -{\bf t}';k^{-1})\frac{dk}{2\pi ik}=0,
\quad x-x'\in \eta \ZZ .
\eeq
We can represent the wave functions in the form
\beq\label{tau1}
\begin{array}{l}
\psi (x, {\bf t}, -{\bf t};k)=k^{x/\eta}e^{\xi ({\bf t}, k)}w(x, {\bf t}; k), \quad k\to \infty ,
\\ \\
\psi (x, {\bf t}, -{\bf t};k^{-1})=k^{-x/\eta}e^{-\xi ({\bf t}, k)}
\bar w(x, {\bf t}; k), \quad k\to \infty ,
\end{array}
\eeq
then the bilinear relation can be written as
\beq\label{tau3}
\begin{array}{c}
\displaystyle{
\oint_{C_{\infty}}k^{n-1}e^{\xi ({\bf t}-{\bf t}', k)}w(x, {\bf t}; k)
\bar w(x\! -\! n\eta , {\bf t}'; k)dk}
\\ \\
=\displaystyle{
\oint_{C_{0}}k^{n-1}e^{-\xi ({\bf t}-{\bf t}', k^{-1})}\bar w(x, {\bf t}; k^{-1})
w(x\! -\! n\eta , {\bf t}'; k^{-1})dk}.
\end{array}
\eeq
One always can normalize the functions $w(x, {\bf t}; k)$, $\bar w(x, {\bf t}; k)$
in the following way:
\beq\label{tau2}
w(x, {\bf t}; \infty )=1, \quad \bar w(x, {\bf t}; \infty )=r(x, {\bf t})
=e^{\varphi (x, {\bf t})}.
\eeq

Now, choosing ${\bf t}-{\bf t}'$ and $n$ in some special ways, one is able to obtain
different relations for the functions $w(x, {\bf t}; k)$, $\bar w(x, {\bf t}; k)$
with certain shifts of the variables.

\paragraph{1. ${\bf t}-{\bf t}'=[a^{-1}]$, $n=1$.} In this case
$\displaystyle{e^{\xi ({\bf t}-{\bf t}', k)}=\frac{a}{a-k}}$ and the bilinear relation
acquires the form
$$
\begin{array}{c}
\displaystyle{
\oint_{C_{\infty}}\frac{a}{a-k}w(x, {\bf t}; k)
\bar w(x\! -\! \eta , {\bf t}-[a^{-1}]; k)dk}
\\ \\
=\displaystyle{
\oint_{C_{0}}\Bigl (1-\frac{1}{ka}\Bigr )
\bar w(x, {\bf t}; k^{-1})
w(x\! -\! \eta , {\bf t}-[a^{-1}]; k^{-1})dk}.
\end{array}
$$
The residue calculus yields
\beq\label{tau4}
w(x, {\bf t}; a)\bar w(x-\eta , {\bf t}-[a^{-1}]; a)=r(x-\eta , {\bf t}-[a^{-1}])-a^{-2}
r(x, {\bf t}).
\eeq

\paragraph{2. ${\bf t}-{\bf t}'=[a^{-1}]+[b^{-1}]$, $n=2$.} In this case
the bilinear relation
acquires the form
$$
\begin{array}{c}
\displaystyle{
\oint_{C_{\infty}}\frac{abk}{(a-k)(b-k)}w(x, {\bf t}; k)
\bar w(x\! -\! 2\eta , {\bf t}-[a^{-1}]-[b^{-1}]; k)dk}
\\ \\
=\displaystyle{
\oint_{C_{0}}k\Bigl (1-\frac{1}{ka}\Bigr )\Bigl (1-\frac{1}{kb}\Bigr )
\bar w(x, {\bf t}; k^{-1})
w(x\! -\! 2\eta , {\bf t}-[a^{-1}]-[b^{-1}]; k^{-1})dk}.
\end{array}
$$
The residue calculus yields
\beq\label{tau5}
\begin{array}{c}
\displaystyle{
\frac{ab}{a\! -\! b}\Bigl (
aw(x, {\bf t};a)\bar w(x\! -\! 2\eta , {\bf t}\! -\! [a^{-1}]\! -\! [b^{-1}]; a)\! - \!
bw(x, {\bf t};b)\bar w(x\! -\! 2\eta , {\bf t}\! -\! [a^{-1}]\! -\! [b^{-1}]; b)\Bigr )}
\\ \\
=ab r(x-2\eta , {\bf t}-[a^{-1}]-[b^{-1}])-(ab)^{-1} r(x, {\bf t}).
\end{array}
\eeq

\paragraph{3. ${\bf t}-{\bf t}'=[a^{-1}]-[b^{-1}]$, $n=0$.} In this case
$$
\begin{array}{c}
\displaystyle{
\oint_{C_{\infty}}k^{-1}\, \frac{a(b-k)}{b(a-k)}w(x, {\bf t}; k)
\bar w(x , {\bf t}-[a^{-1}]+[b^{-1}]; k)dk}
\\ \\
=\displaystyle{
\oint_{C_{0}}k^{-1}\, \frac{k-a^{-1}}{k-b^{-1}}
\bar w(x, {\bf t}; k^{-1})
w(x, {\bf t}-[a^{-1}]+[b^{-1}]; k^{-1})dk}
\end{array}
$$
and residue calculus yields
\beq\label{tau6}
\begin{array}{c}
\displaystyle{
\Bigl (1-\frac{a}{b}\Bigr )
w(x, {\bf t};a)\bar w(x, {\bf t}\! -\! [a^{-1}]\! +\! [b^{-1}]; a)\! - \!
\Bigl (1-\frac{b}{a}\Bigr )
\bar w(x, {\bf t};b) w(x, {\bf t}\! -\! [a^{-1}]\! +\! [b^{-1}]; b)}
\\ \\
\displaystyle{=\frac{b}{a}\, r(x, {\bf t})-\frac{a}{b}\,
r(x, {\bf t}-[a^{-1}]+[b^{-1}])}.
\end{array}
\eeq

Expressing $\bar w$ through $w$ with the help of (\ref{tau4}), we can represent
the other two relations, (\ref{tau5}) and (\ref{tau6}), as a system of two equations
for two ``variables''
\beq\label{tau7}
X_a=\frac{w(x-\eta , {\bf t}-[b^{-1}];a)}{w(x, {\bf t};a)}, \quad
X_b=\frac{w(x-\eta , {\bf t}-[a^{-1}];b)}{w(x, {\bf t};b)}.
\eeq
The system has the form
\beq\label{tau8}
\left \{
\begin{array}{l}
\displaystyle{
\begin{array}{l}\displaystyle{\frac{ab}{a\! -\! b}\Bigl [
ag(x\! -\! \eta , {\bf t}\! -\! [b^{-1}];a)X_a^{-1}-
bg(x\! -\! \eta , {\bf t}\! -\! [a^{-1}];b)X_b^{-1}\Bigr ]}
\\ \\
\phantom{aaaaaaaaaaaaaaaaaaaaaaaaaa}\displaystyle{=
ab r(x\! -\! 2\eta , {\bf t}\! -\! [a^{-1}]\! -\! [b^{-1}])-(ab)^{-1}r(x, {\bf t})}
\end{array}}
\\ \\
\displaystyle{
\Bigl (1\! -\! \frac{a}{b}\Bigr )g(x, {\bf t};a)X_a -
\Bigl (1\! -\! \frac{b}{a}\Bigr )g(x, {\bf t};b)X_b =
\frac{b}{a}r(x\! -\! \eta , {\bf t}\! -\! [b^{-1}])-
\frac{a}{b}r(x\! -\! \eta , {\bf t}\! -\! [a^{-1}]),}
\end{array}
\right.
\eeq
where
\beq\label{tau9}
g(x, {\bf t};z)=r(x-\eta , {\bf t}-[z^{-1}])-z^{-2}r(x, {\bf t}).
\eeq
Next, we take the product of the left hand sides of the two equations (\ref{tau8})
and equate it 
to the product of the right hand sides.
After some transformations, we obtain the remarkable relation
\beq\label{tau10}
\left (\frac{X_a}{X_b}\right )^2=
\frac{w^2 (x, {\bf t};b) w^2(x-\eta ,
{\bf t}-[b^{-1}];a)}{w^2 (x, {\bf t};a) w^2(x-\eta , {\bf t}-[a^{-1}];b)}=
\frac{g (x, {\bf t};b) g(x-\eta ,
{\bf t}-[b^{-1}];a)}{g (x, {\bf t};a) g(x-\eta , {\bf t}-[a^{-1}];b)}
\eeq
which implies that
$$
w_0(x, {\bf t};z):=w(x, {\bf t};z)g^{-1/2}(x, {\bf t};z)
$$
obeys the relation
\beq\label{tau11}
\frac{w_0 (x, {\bf t};b) w_0(x-\eta ,
{\bf t}-[b^{-1}];a)}{w_0 (x, {\bf t};a) w_0(x-\eta , {\bf t}-[a^{-1}];b)}=1.
\eeq
It follows from this relation that there exists a function $\tau^C (x, {\bf t})$ such that
\beq\label{tau12}
w_0(x, {\bf t};z)=\frac{\tau^C(x-\eta , {\bf t}-[z^{-1}])}{\tau^C(x, {\bf t})}.
\eeq
The proof is almost literally a repetition of the proof of a similar statement
for the CKP hierarchy presented in \cite{KZ21}.

Therefore, we have
\beq\label{tau12a}
w(x, {\bf t};k)=g^{1/2}(x, {\bf t};k)
\frac{\tau^C(x-\eta , {\bf t}-[k^{-1}])}{\tau^C(x, {\bf t})}
\eeq
with $g(x, {\bf t};k)$ as in (\ref{tau9}).
The normalization of $w$ implies that
$$
1=w(x, {\bf t}, \infty )=r^{1/2}(x-\eta , {\bf t})
\frac{\tau^C(x-\eta , {\bf t})}{\tau^C(x, {\bf t})},
$$
whence
\beq\label{tau13}
r(x, {\bf t})=\left (\frac{\tau^C(x+\eta , {\bf t})}{\tau^C (x, {\bf t})}\right )^2.
\eeq
On the other hand, we know that
\beq\label{tau14}
r(x, {\bf t})=\frac{\tau^T(x+\eta , {\bf t}, -{\bf t})}{\tau^T (x, {\bf t}, -{\bf t})},
\eeq
where $\tau^T$ is the tau-function of the 2D Toda lattice hierarchy. This implies the
following relation between the two tau-functions:
\beq\label{tau15}
\tau^T (x, {\bf t}, -{\bf t})=C({\bf t})(\tau^C (x, {\bf t}))^2,
\eeq
where $C({\bf t})$ is a quasi-constant in $x$ (i.e., 
it is an $\eta$-periodic function of $x$) depending on ${\bf t}$. 
Below we shall see
that in fact $C$ does not depend on ${\bf t}$.

Finally, we conclude that the factor
$w(x, {\bf t};k)$ which enters the $k\to \infty$ asymptotics of the
wave function $\psi (x, {\bf t}, -{\bf t};k)$ (see (\ref{tau1}))
is expressed through the tau-function as follows:
\beq\label{tau16}
w(x, {\bf t};k)=
\left [1 - k^{-2}
\left (\frac{\tau^C(x+\eta , {\bf t})\tau^C (x-\eta , {\bf t}-[k^{-1}])}{\tau^C
(x, {\bf t})\tau^C (x, {\bf t}\! -\! [k^{-1}])}\right )^2 \right ]^{1/2}
\! \frac{\tau^C (x, {\bf t}-[k^{-1}])}{\tau^C (x, {\bf t})}.
\eeq
The function $\bar w(x, {\bf t};k)$ which enters the $k\to 0$ asymptotics of the
function $\psi (x, {\bf t}, -{\bf t};k)$ can be found from the relation
(\ref{tau4}) which reads
$$
w(x, {\bf t}; k)\bar w(x-\eta , {\bf t}-[k^{-1}]; k)=g(x, {\bf t};k).
$$
After a simple algebra, we obtain:
\beq\label{tau17}
\begin{array}{l}
\displaystyle{
\bar w(x, {\bf t};k)=
\left [1 - k^{-2}
\left (\frac{\tau^C(x, {\bf t})\tau^C (x+2\eta , {\bf t}+[k^{-1}])}{\tau^C
(x+\eta , {\bf t})\tau^C (x+\eta , {\bf t}\! +\! [k^{-1}])}\right )^2 \right ]^{1/2}}
\\ \\
\displaystyle{\phantom{aaaaaaaaaaaaaaa}
\times \frac{\tau^C (x+\eta , {\bf  t})}{\tau^C (x, {\bf t})}\,
\frac{\tau^C (x+\eta , {\bf t}+[k^{-1}])}{\tau^C (x, {\bf t})}}.
\end{array}
\eeq

We can represent equations (\ref{tau16}), (\ref{tau17}) in a more suggestive form.
Introduce modified ``wave functions'' $\chi$, $\bar \chi$ which are connected with
$\tau^C$ in the same way as $\psi$, $\psi^*$ are connected with $\tau^T$:
\beq\label{chi1}
\chi (x, {\bf t};k)=k^{x/\eta}e^{\xi ({\bf t}, k) -\frac{1}{2}\varphi (x, {\bf t})}
\frac{\tau^C (x, {\bf t}-[k^{-1}])}{\tau^C (x, {\bf t})},
\eeq
\beq\label{chi2}
\bar \chi (x, {\bf t};k)=k^{-x/\eta}e^{-\xi ({\bf t}, k)}
\frac{\tau^C (x+\eta , {\bf t}+[k^{-1}])}{\tau^C (x, {\bf t})},
\eeq
where
\beq\label{chi2b}
e^{\varphi (x, {\bf t})}=r(x, {\bf t}) =
\left (\frac{\tau^C(x+\eta , {\bf t})}{\tau^C (x, {\bf t})}\right )^2.
\eeq
Recalling (\ref{tau13}),
it is easy to check that formulas (\ref{tau16}), (\ref{tau17}) are equivalent to
\beq\label{chi4}
\psi (x, {\bf t}; k)=e^{\frac{1}{2}\varphi (x, {\bf t})}
\sqrt{\chi^2(x, {\bf t};k)-\chi^2(x\! -\! \eta , {\bf t};k)}, \quad k\to \infty ,
\eeq
\beq\label{chi4a}
\psi (x, {\bf t}; k^{-1})=e^{\frac{1}{2}\varphi (x, {\bf t})}
\sqrt{\bar \chi^2(x, {\bf t};k)-\bar \chi^2(x\! +\! \eta , {\bf t};k)}, 
\quad k\to \infty .
\eeq
These formulas resemble the corresponding formula for the CKP hierarchy (see \cite{KZ21}),
with the $x$-derivative substituted by the difference.
\square

We already proved relation (\ref{tau15}) between $\tau^C$ and $\tau^T$. 
Now we are going to
prove that $C({\bf t})=C$ is a quasi-constant in $x$ 
which does not depend on the times, so that
$\tau^C$ is essentially the square root of $\tau^T$ (restricted to the submanifold
${\bf t}+\bar {\bf t}=0$ and satisfying the ``turning points'' condition (\ref{inv17a})).

\begin{theorem}\label{theorem:square}
The tau-functions $\tau^C$ and $\tau^T$ are 
related as $\tau^T =C (\tau^C)^2$, where $C$ is
a quasi-constant in $x$, i.e., 
the tau-function of the C-Toda hierarchy is essentially square root
of the 2D Toda lattice tau-function.
\end{theorem}

\noindent
{\it Proof.}
First of all, we recall that together with (\ref{tau16}) alternative formulas for
$w(x, {\bf t};k)$ through $\tau^T$ hold:
\beq\label{tau18}
w(x, {\bf t};k)=\frac{\tau^T(x, {\bf t}-[k^{-1}], -{\bf t})}{\tau^T (x, {\bf t}, -{\bf t})}=
\frac{\tau^T(x, {\bf t}, -{\bf t}+[k^{-1}])}{\tau^T (x, {\bf t}, -{\bf t})}
\eeq
(the second equality is due to (\ref{inv16})). Substituting them into (\ref{tau16}) and
taking square of both sides, we obtain the relation
$$
\left (\frac{\tau^C(x, {\bf t}-[k^{-1}])}{\tau^C (x, {\bf t})}\right )^2
-k^{-2}\left ( \frac{\tau^C (x+\eta ,
{\bf t})\tau^C(x-\eta , {\bf t}-[k^{-1}])}{\tau^C (x, {\bf t})\tau^C (x, {\bf t})}
\right )^2
$$
$$
=\frac{\tau^T(x+\eta , {\bf t}, -{\bf t})}{\tau^T(x, {\bf t}, -{\bf t})}\,
\left [\frac{\tau^T (x, {\bf t}-[k^{-1}], -{\bf t})
\tau^T (x, {\bf t}, -{\bf t}+[k^{-1}])}{\tau^T(x+\eta , {\bf t}, -{\bf t})
\tau^T(x, {\bf t}, -{\bf t})}\right ].
$$
Now we are going to use the Hirota-Miwa equation (\ref{bil2}) 
for $a=b=k$ which we rewrite
in the form
$$
\frac{\tau^T (x, {\bf t}-[k^{-1}], -{\bf t})
\tau^T (x, {\bf t}, -{\bf t}+[k^{-1}])}{\tau^T(x+\eta , {\bf t}, -{\bf t})
\tau^T(x, {\bf t}, -{\bf t})}
$$
$$=
\frac{\tau^T(x, {\bf t}-[k^{-1}], -{\bf t}+[k^{-1}])}{\tau^T
(x+\eta , {\bf t}, -{\bf t})}-
k^{-2}
\frac{\tau^T(x-\eta , {\bf t}-[k^{-1}], -{\bf t}+[k^{-1}])}{\tau^T
(x+\eta , {\bf t}, -{\bf t})}.
$$
Substituting the right hand side instead of the brackets $[\ldots ]$ in the previous relation,
we get
$$
\left (\frac{\tau^C(x, {\bf t}-[k^{-1}])}{\tau^C (x, {\bf t})}\right )^2
-k^{-2}\left ( \frac{\tau^C (x+\eta ,
{\bf t})\tau^C(x-\eta , {\bf t}-[k^{-1}])}{\tau^C (x, {\bf t})\tau^C (x, {\bf t})}
\right )^2
$$
$$
=
\frac{\tau^T(x, {\bf t}-[k^{-1}], -{\bf t}+[k^{-1}])}{\tau^T(x, {\bf t}, -{\bf t})}-
k^{-2}
\frac{\tau^T(x-\eta , {\bf t}-[k^{-1}], -{\bf t}+[k^{-1}])
\tau^T(x+\eta , {\bf t}, -{\bf t})}{\tau^T(x, {\bf t}, -{\bf t})\tau^T(x, {\bf t}, -{\bf t})}.
$$
Plugging here (\ref{tau15}), we obtain
$$
\left (\frac{C({\bf t}-[k^{-1}])}{C({\bf t})}-1\right )
\left [\left (\frac{\tau^C (x, {\bf t}-[k^{-1}])}{\tau^C(x+\eta , {\bf t})}\right )^2-
k^{-2}\left (\frac{\tau^C (x-\eta , {\bf t}-[k^{-1}])}{\tau^C(x, {\bf t})}\right )^2\right ]=0.
$$
Since the factor in the square brackets is nonzero, we conclude that
$C({\bf t}-[k^{-1}])-C({\bf t})\equiv 0$ as a power series in $k$.
This implies that $C({\bf t})$ does not depend on ${\bf t}$ and, therefore,
$\tau^C =\sqrt{\tau^T}$.
\square

\section{Turning points of Ruijsenaars-Schneider model}

\subsection{Elliptic Ruijsenaars-Schneider model}

Here we collect the main facts on the elliptic Ruijsenaars-Schneider system \cite{RS86}
following the paper \cite{Ruij87}.

The $N$-particle elliptic Ruijsenaars-Schneider system
(a relativistic extension of the Calogero-Moser system) is a completely integrable model.
The canonical Poissson brackets between coordinates and momenta are
$\{x_i, p_j\}=\delta_{ij}$.
The integrals of motion in involution have the form
\beq\label{intr1}
I_k= \sum_{I\subset \{1, \ldots , N\}, \, |I|=k}
\exp \Bigl (\sum_{i\in I}p_i\Bigr ) \prod_{i\in I, j\notin I}\frac{\sigma
(x_i-x_j+\eta )}{\sigma (x_i-x_j)}, \quad k=1, \ldots , N,
\eeq
where $\sigma (x)$ is the Weierstrass $\sigma$-function and
$\eta$ is a parameter which has a meaning of the inverse velocity of light.
The $\sigma$-function
with quasi-periods $2\omega_1$, $2\omega_2$ such that
${\rm Im} (\omega_2/ \omega_1 )>0$
is defined as
$$
\sigma (x)=\sigma (x |\, \omega_1 , \omega_2)=
x\prod_{s\neq 0}\Bigl (1-\frac{x}{s}\Bigr )\, e^{\frac{x}{s}+\frac{x^2}{2s^2}},
\quad \! s=2\omega_1 m_1+2\omega_2 m_2 \quad \mbox{with integer $m_1, m_2$}.
$$
It is connected with the Weierstrass
$\zeta$- and $\wp$-functions by the formulas $\zeta (x)=\sigma '(x)/\sigma (x)$,
$\wp (x)=-\zeta '(x)=-\p_x^2\log \sigma (x)$.
Important particular cases of (\ref{intr1}) are
\beq\label{intr2}
I_1= H_1=\sum_i e^{p_i}\prod_{j\neq i} \frac{\sigma
(x_i-x_j+\eta )}{\sigma (x_i-x_j)}
\eeq
which is the Hamiltonian $H_1$ of the chiral Ruijsenaars-Schneider model and
\beq\label{intr2a}
I_N=\exp \Bigl (\sum_{i=1}^{N}p_i\Bigr ).
\eeq
It is natural to put $I_0=1$.
Comparing to the paper \cite{Ruij87}, our formulas differ by the canonical transformation
$$
e^{p_i}\to e^{p_i}\prod_{j\neq i}\left ( \frac{\sigma
(x_i-x_j+\eta )}{\sigma (x_i-x_j-\eta )}\right )^{1/2}, \quad x_i\to x_i,
$$
which allows one to eliminate square roots in \cite{Ruij87}.

Let us denote the time variable of the Hamiltonian flow with the Hamiltonian $H=I_1$
by $t_1$.
The velocities of the particles are
\beq\label{intr4}
\dot x_i =\frac{\p H_1}{\p p_i}=e^{p_i}\prod_{j\neq i} \frac{\sigma
(x_i-x_j+\eta )}{\sigma (x_i-x_j)},
\eeq
where dot means the $t_1$-derivative.
The Hamiltonian equations $\dot p_i=-\p H_1/ \p x_i$ are equivalent to the following
equations of motion:
\beq\label{te4}
\begin{array}{lll}
\ddot x_i &=&\displaystyle{-\sum_{k\neq i}\dot x_i\dot x_k \Bigl (
\zeta (x_i-x_k+\eta )+\zeta (x_i-x_k-\eta )-2\zeta (x_i-x_k)\Bigr )}
\\ && \\
&=&\displaystyle{\sum_{k\neq i}\dot x_i\dot x_k\frac{\wp '(x_i-x_k)}{\wp (\eta )-
\wp (x_i-x_k)}.}
\end{array}
\eeq

One can also introduce integrals of motion $I_{-k}$ as
\beq\label{intr1b}
I_{-k}=I_{N}^{-1}I_{N-k}=\sum_{I\subset \{1, \ldots , N\}, \, |I|=k}
\exp \Bigl (-\sum_{i\in I}p_i\Bigr ) \prod_{i\in I, j\notin I}\frac{\sigma
(x_i-x_j-\eta )}{\sigma (x_i-x_j)}.
\eeq
In particular,
\beq\label{intr2b}
I_{-1}= \sum_i e^{-p_i}\prod_{j\neq i} \frac{\sigma
(x_i-x_j-\eta )}{\sigma (x_i-x_j)}.
\eeq
It can be easily verified that equations of motion
in the time $\bar t_1$ corresponding to the Hamiltonian $\bar H_1=\sigma^2(\eta )I_{-1}$
are the same
as (\ref{te4}):
\beq\label{intr5}
\stackrel{\circ \circ}{x}_i=\displaystyle{-\sum_{k\neq i}\stackrel{\circ}{x}_i \,
\stackrel{\circ}{x}_k \Bigl (
\zeta (x_i-x_k+\eta )+\zeta (x_i-x_k-\eta )-2\zeta (x_i-x_k)\Bigr )}.
\eeq
Here and below $\circ$ means the $\bar t_1$-derivative. The velocity $\stackrel{\circ}{x}_i$
is given by
\beq\label{vel1}
\stackrel{\circ}{x}_i =\frac{\p \bar H_1}{\p p_i}=
-\sigma^2(\eta )e^{-p_i}\prod_{j\neq i} \frac{\sigma
(x_i-x_j-\eta )}{\sigma (x_i-x_j)}.
\eeq
Multiplying (\ref{intr4}) and (\ref{vel1}), we obtain the important
relation between
$\dot x_i$ and $\stackrel{\circ}{x}_i$:
\beq\label{intr6}
\dot x_i \! \stackrel{\circ}{x}_i=-\sigma ^2(\eta )
\prod_{k\neq i}\frac{\sigma (x_i-x_k+\eta )\sigma (x_i-x_k-\eta )}{\sigma ^2 (x_i-x_k)}
\eeq
(see \cite{KZ95,PZ21}). The physical Hamiltonian of the Ruijsenaars-Schneider model
is $H=H_1 +\bar H_1$.

\subsection{The Ruijsenaars-Schneider model from the 2D Toda lattice}

In the paper \cite{KZ95} (see also the review \cite{Z19}) it was shown that the Ruijsenaars-Schneider
dynamics is the same as dynamics of poles of elliptic solutions to the 2D Toda equation
in the Toda times $t_1$, $\bar t_1$. Later, in \cite{PZ21}, this observation
was extended to a complete isomorphism between the elliptic
Ruijsenaars-Schneider model (with higher
Hamiltonian flows) and elliptic solutions to
the whole 2D Toda lattice hierarchy.

In terms of the tau-function, the 2D Toda equation (the first equation of the hierarchy)
reads
\beq\label{toda}
\p_t \p_{\bar t}\log \tau (x)=-\frac{\tau (x+\eta )\tau (x-\eta )}{\tau^2(x)},
\eeq
where $t=t_1$, $\bar t =\bar t_1$.
The tau-function for elliptic solutions of the 2D Toda lattice hierarchy
has the form
\beq\label{tauell}
\tau (x, {\bf t}, \bar {\bf t})=\exp \Bigl (-\sum_{k\geq 1}kt_k \bar t_k \Bigr )
\prod_{i=1}^{N}\sigma (x-x_i({\bf t}, \bar {\bf t})).
\eeq
The zeros $x_i$ of the tau-function are poles of the
solution. They are assumed to be
all distinct.

One can see that the relation
(\ref{intr6}) is a consequence of the 2D Toda equation. Indeed, (\ref{intr6}) is obtained
from (\ref{toda}) with the tau-function (\ref{tauell}) by equating the coefficients at the highest
(second order) poles at $x=x_i$ of both sides.

\subsection{The Lax matrix and the spectral curve}

The equations of motion of the Ruijsenaars-Schneider model admit the Lax representation.
The Lax matrix depends on a spectral parameter $\lambda$ and has the form \cite{KZ95,PZ21}
\beq\label{lax1}
L_{ij}(\lambda )=e^{-(x_i-x_j)\zeta (\lambda )}\,\dot x_i \,
\frac{\sigma (x_i-x_j-\eta +\lambda )}{\sigma (\lambda )\sigma (x_i-x_j-\eta )},
\quad i,j=1, \ldots , N.
\eeq
The characteristic polynomial of the Lax matrix is the generating function of the integrals
of motion (\ref{intr1}):
\beq\label{lax2}
\det \Bigl (zI-L(\lambda )\Bigr )=\sum_{n=0}^N
\frac{\sigma (\lambda -n\eta )}{\sigma (\lambda )\sigma^n (\eta )}\, I_n z^{N-n}
\eeq
(here $I$ is the unity matrix).

The characteristic equation
\beq\label{lax3}
R(z, \lambda ):=\det \Bigl (zI-L(\lambda )\Bigr )=0
\eeq
defines a Riemann surface $\tilde \Gamma$ which is 
an $N$-sheet covering of the $\lambda$-plane. 
Any point of it is
$P=(z, \lambda )$, where $z, \lambda$ are connected by equation (\ref{lax3}). There are $N$
points of the curve above each point $\lambda$. 
It is easy to see from the right hand side of
(\ref{lax2})
that the Riemann surface $\tilde \Gamma$ 
is invariant under the simultaneous transformations
\beq\label{trans}
\lambda \mapsto \lambda +2\omega_{\alpha}, \quad z\mapsto 
e^{-2\zeta (\omega_{\alpha})\eta}z.
\eeq
The factor of $\tilde \Gamma$ over the transformations (\ref{trans}) is an
algebraic curve $\Gamma$ which covers the elliptic curve with periods $2\omega_{\alpha}$.
It is the spectral curve of the Ruijsenaars-Schneider model.
The points $P_{\infty}=(\infty , 0)$ and
$P_0=(0, N\eta )$ are special. They are marked points of the algebraic curve, where
the Baker-Akhiezer function for the elliptic solutions of the 2D Toda lattice hierarchy
has essential singularities.

Let us note that the Lax matrix has the form of the elliptic Cauchy matrix times diagonal
matrices from the left and from the right. The explicit form of determinant of
the elliptic Cauchy
matrix is known:
\beq\label{lax5}
\det_{1\leq i,j\leq N}\left (
\frac{\sigma (x_i-y_j +\lambda )}{\sigma (\lambda )\sigma (x_i-y_j)}\right )
=\frac{\sigma \Bigl (\lambda +\sum\limits_{i=1}^N (x_i-y_i)\Bigr )}{\sigma (\lambda )}
\frac{\prod\limits_{i<j}\sigma (x_i-x_j)\sigma (y_j-y_i)}{\prod\limits_{i,j}\sigma (x_i-y_j)}.
\eeq
This allows one to obtain an explicit expression for the matrix inverse
to the $L(\lambda )$:
\beq\label{lax4}
\begin{array}{l}
\displaystyle{
(L^T(\lambda ))^{-1}_{ij}=e^{(x_i-x_j)\zeta (\lambda )}\,\dot x_i^{-1} \,
\frac{\sigma (x_i-x_j-\eta +N\eta -\lambda )\sigma^2(\eta)}{\sigma (N\eta -\lambda )
\sigma (x_i-x_j-\eta )}}
\\ \\
\phantom{aaaaaaaaaaaaaaaaaaaaa}\displaystyle{\times
\prod_{k\neq i} \frac{\sigma (x_i-x_k-\eta )}{\sigma (x_i-x_k)}
\prod_{m\neq i} \frac{\sigma (x_j-x_m+\eta )}{\sigma (x_j-x_m)}.}
\end{array}
\eeq
Here $L^T$ is the transposed matrix.

\subsection{Turning points}

Turning points of the Ruijsenaars-Schneider model are defined by the conditions
\beq\label{turn1}
\dot x_i + \stackrel{\circ}{x}_i=0 \quad \mbox{or} \quad
(\p_{t_1}+\p_{\bar t_1})x_i =0, \quad i=1, \ldots , N.
\eeq
They mean that the velocities of all particles in the physical Ruijsenaars-Schneider model
with the Hamiltonian $H=H_1+\bar H_1$ are equal to zero.
From equation (\ref{intr6}) we see that this is equivalent to
\beq\label{turn2}
\begin{array}{l}
\displaystyle{
\dot x_i = \sigma (\eta )
\prod_{k\neq i}\frac{(\sigma (x_i-x_k+\eta )\sigma (x_i-x_k-\eta ))^{1/2}}{\sigma (x_i-x_k)}}
\\ \\
\displaystyle{\phantom{aaaaaaaaaaaaaaaaaa}
=\sigma ^N(\eta )\prod_{k\neq i}\sqrt{\wp (\eta )-\wp (x_i-x_k)}}
\end{array}
\eeq
or
\beq\label{turn2a}
e^{p_i}=\sigma (\eta )\prod_{j\neq i}
\left (\frac{\sigma (x_i-x_j-\eta )}{\sigma (x_i-x_j+\eta )}\right )^{1/2}.
\eeq

The turning points form an $N$-dimensional submanifold ${\cal T}\subset {\cal P}$ of the
$2N$-dimensional phase space ${\cal P}$.

\begin{proposition}
The Hamiltonian flow
$\p_{T_1}=\p_{t_1}-\p_{\bar t_1}$ with the Hamiltonian $\bar H=H_1-\bar H_1$
preserves the submanifold ${\cal T}$.
\end{proposition}

\noindent
{\it Proof.}
The corresponding time variable will be denoted as
$T_1=\frac{1}{2}\, (t_1-\bar t_1)$. We have:
\beq\label{pr1}
\stackrel{*}{x}_i =\frac{\p \bar H}{\p p_i}=2\sigma (\eta )
\prod_{k\neq i}\frac{(\sigma (x_i-x_k+\eta )\sigma (x_i-x_k-\eta ))^{1/2}}{\sigma (x_i-x_k)}
\quad \mbox{on ${\cal T}$},
\eeq
where star means the $T_1$-derivative. Taking the $T_1$-derivative of (\ref{turn2a}),
we get
\beq\label{pr2}
\stackrel{*}{p}_i=\frac{1}{2}\sum_{j\neq i}(\stackrel{*}{x}_i - \stackrel{*}{x}_j)
\Bigl ( \zeta (x_i-x_j-\eta )-\zeta (x_i-x_j+\eta )\Bigr ).
\eeq
At the same time,
$$
\stackrel{*}{p}_i=-\frac{\p \bar H}{\p x_i}=
-e^{p_i}
\prod_{j\neq i}\frac{\sigma (x_i-x_j +\eta )}{\sigma (x_i-x_j)}
\sum_{l\neq i} \Bigl ( \zeta (x_i-x_l+\eta )-\zeta (x_i-x_l)\Bigr )
$$
$$
+\sigma^2(\eta )e^{-p_i}\prod_{j\neq i}\frac{\sigma (x_i-x_j -\eta )}{\sigma (x_i-x_j)}
\sum_{l\neq i} \Bigl ( \zeta (x_i-x_l-\eta )-\zeta (x_i-x_l)\Bigr )
$$
$$
+\sum_{l\neq i}e^{p_l}\prod_{j\neq l}\frac{\sigma (x_l-x_j +\eta )}{\sigma (x_l-x_j)}
\Bigl ( \zeta (x_l-x_i+\eta )-\zeta (x_l-x_i)\Bigr )
$$
$$
-\sigma^2(\eta )
\sum_{l\neq i}e^{-p_l}\prod_{j\neq l}\frac{\sigma (x_l-x_j -\eta )}{\sigma (x_l-x_j)}
\Bigl ( \zeta (x_l-x_i-\eta )-\zeta (x_l-x_i)\Bigr ).
$$
Plugging here the turning point condition (\ref{turn2a}) and using (\ref{pr1}), we
obtain (\ref{pr2}). This means that the submanifold ${\cal T}$ is indeed invariant
under the $T_1$-flow.
\square

Now we are going to prove that for any turning point the spectral curve $\Gamma$ admits
a holomorphic involution.

\begin{theorem}\label{theorem:involution}
For any turning point the spectral curve $\Gamma$ admits
the holomorphic involution
\beq\label{turn5}
\iota : (z, \lambda )\to (z^{-1}, N\eta \! -\! \lambda ).
\eeq
\end{theorem}

\noindent
{\it Proof.}
Substituting (\ref{turn2}) into (\ref{lax1}) and
(\ref{lax4}), we see that
\beq\label{turn3}
(L^T(\lambda ))^{-1}=UL(N\eta -\lambda )U^{-1},
\eeq
where $U=\mbox{diag} (U_1, \ldots , U_N)$ is the diagonal matrix with
\beq\label{turn4}
U_i=e^{x_i(\zeta (\lambda )+\zeta (N\eta \! -\! \lambda ))}\prod_{k\neq i}
\frac{\sigma (x_i-x_k)}{\sigma (x_i-x_k+\eta )}.
\eeq
Therefore, the spectral curve (\ref{lax3}) has the holomorphic involution (\ref{turn5}).
\square

Note that the involution interchanges the two marked points: $\iota P_{\infty}=P_0$, $\iota P_0=P_{\infty}$. The following proposition characterizes fixed points of the involution.

\begin{proposition}
The involution $\iota$ has 2 fixed points for even $N$ and 4 fixed points for odd $N$.
\end{proposition}

\noindent
{\it Proof.} The fixed points may lie above points $\lambda_{*}$ such that
$\lambda_{*} =N\eta -\lambda_{*}$ modulo the lattice with periods $2\omega_{\alpha}$, i.e.
$\lambda_{*}=\frac{1}{2}N\eta -\omega$, where $\omega$ is either $0$ or one of the three
half-periods. Substituting this into the equation of the spectral curve (\ref{lax2})
and taking into account that for turning points it holds $I_k=I_{N-k}$, we conclude that
for even $N$ the fixed points are $(\pm 1, \frac{1}{2}N\eta )$ while for odd $N$
the fixed points are $(1, \frac{1}{2}N\eta )$ and three points
$(-e^{-\zeta (\omega )\eta}, \frac{1}{2}N\eta -\omega )$ for the three half-periods $\omega$.
\square

We have shown that from the condition on the turning points it follows that the spectral
curve has a holomorphic involution with fixed points. 
Now we are going to prove the inverse statement:
the involution of the curve (which can be not necessarily the spectral curve 
of the Ruijsenaars-Schneider model) having fixed points 
implies the turning points condition for zeros of the tau-function corresponding 
to the algebraic-geometrical solution constructed from the curve
according to the general construction of quasi-periodic (algebraic-geometrical) solutions
\cite{Krichever77,Krichever77a}.
Quasi-periodic solutions to the Toda lattice equation were constructed
in \cite{Krichever81}. The algebraic-geometrical data include an algebraic curve
$\Gamma$ of genus $g$
with two marked points $P_0$, $P_{\infty}$, local parameters near the marked points and
an effective divisor ${\cal D}$ of degree $g$ on $\Gamma$. Algebraic-geometrical
solutions of the constrained Toda hierarchy were recently constructed in \cite{kr-inv}.

\begin{theorem}\label{theorem:curve}
Let $\Gamma$ be an algebraic curve with holomorphic involution $\iota$
which has fixed points and two marked points $P_{\infty}$, $P_0$ such that
$P_0 =\iota P_{\infty}$.
Let $k^{-1}$ be a local parameter in the vicinity of $P_{\infty}$ ($k^{-1}(P_{\infty})=0$),
we assume that the local parameter in the vicinity of $P_0$ is $k$ ($k(P_0)=0$), so that
$\iota (k)=k^{-1}$. Besides, we fix an effective divisor
${\cal D}$ of degree $g$ on $\Gamma$ such that
\beq\label{div}
{\cal D}+\iota {\cal D}={\cal K}+P_0+P_{\infty},
\eeq
where ${\cal K}$ is the canonical class.
Then zeros of the tau-function of the solution to the 2D Toda lattice
constructed from these algebraic-geometrical data satisfy the turning points condition.
\end{theorem}

\noindent
{\it Proof.}
Let $\psi (x; P)=\psi (x, t, \bar t;P)$ be the
Baker-Akhiezer function on the curve
$\Gamma$ ($P$ is a point on $\Gamma$). It has simple poles at the points of the divisor
${\cal D}$.
Its behavior in the vicinity of the marked points is
\beq\label{inv1}
\psi (x;P)=\left \{
\begin{array}{l}
\displaystyle{ k^{x/\eta}e^{kt}\Bigl (1+\sum_{s\geq 1} \xi_s (x)k^{-s}\Bigr )},
\quad P\to P_{\infty} \quad (k\to \infty ),
\\ \\
\displaystyle{ e^{\varphi (x)}
k^{x/\eta}e^{k^{-1}\bar t}\Bigl (1+\sum_{s\geq 1} \chi_s (x)k^{s}\Bigr )},
\quad P\to P_{0} \quad (k\to 0).
\end{array}
\right.
\eeq
The function $\varphi (x)$ is expressed through the tau-function as in (\ref{inv2}).
The Baker-Akhiezer function satisfies the linear equation
\beq\label{inv3a}
\p_t \psi (x;P)=\psi (x+\eta ;P)+v(x)\psi (x;P),
\eeq
where
\beq\label{inv4}
v(x)=\p_t \log \frac{\tau (x+\eta)}{\tau (x)}=\dot \varphi (x).
\eeq
Substituting (\ref{inv1}) into (\ref{inv3}), we obtain, in the limit $k\to \infty$:
\beq\label{inv5}
v(x)=\xi_1(x)-\xi_1(x+\eta ), \quad \xi_1(x)=-\p_t \log \tau (x).
\eeq

The dual Baker-Akhiezer function $\psi^*(x;P)$ satisfies the equation
\beq\label{inv3b}
-\p_t \psi^* (x;P)=\psi^* (x-\eta ;P)+v(x)\psi^* (x;P).
\eeq
Its behavior in the vicinity of the marked points is
\beq\label{inv1a}
\psi^* (x;P)=\left \{
\begin{array}{l}
\displaystyle{ k^{-x/\eta}e^{-kt}\Bigl (1+\sum_{s\geq 1} \xi^*_s (x)k^{-s}\Bigr )},
\quad P\to P_{\infty} \quad (k\to \infty ),
\\ \\
\displaystyle{ e^{-\varphi (x)}
k^{-x/\eta}e^{-k^{-1}\bar t}\Bigl (1+\sum_{s\geq 1} \chi^*_s (x)k^{s}\Bigr )},
\quad P\to P_{0} \quad (k\to 0).
\end{array}
\right.
\eeq
Substituting (\ref{inv1a}) into (\ref{inv3a}), we obtain $v(x)=\xi^*_1(x)-\xi^*_1(x+\eta )$.
Comparing with (\ref{inv5}), we conclude that
\beq\label{inv6}
\xi_1^*(x)=-\xi_1(x+\eta).
\eeq

On the curve with involution such that $P_0=\iota P_{\infty}$, we can consider the function
\beq\label{inv7}
\psi^{\iota}(x;P)=\psi (x;\iota P).
\eeq
The condition (\ref{div}) imposed on the divisor ${\cal D}$ and
the behavior of $\psi^{\iota}$
near the marked points imply (due to uniqueness of the Baker-Akhiezer
function) that we can identify
\beq\label{inv8}
\left. \phantom{\int}
\psi^*(x, t, \bar t;P)=e^{-\varphi (x)}\psi^{\iota}(x, t, \bar t;P)\right |_{t+\bar t=0},
\eeq
whence
\beq\label{inv9}
\chi_s (x)= \xi_s^*(x)
\eeq
and the behavior of the function $\psi^{\iota}$ near $P_{\infty}$ is
\beq\label{inv10}
\psi^{\iota}(x;P)=e^{\varphi (x)}
k^{-x/\eta}e^{k^{}\bar t}\Bigl (1+\sum_{s\geq 1} \xi^*_s (x)k^{-s}\Bigr ),
\quad k\to \infty .
\eeq
Substituting this into the linear equation (\ref{inv3}) as $k\to \infty$, we obtain, in the
order $k^{-1}$:
\beq\label{inv11}
\dot \xi_1^*(x)=e^{\varphi (x+\eta )-\varphi (x)}.
\eeq
Equation (\ref{inv6}) allows one to rewrite this relation as
\beq\label{inv12}
\dot \xi_1(x+\eta )=-e^{\varphi (x+\eta )-\varphi (x)},
\eeq
or, using (\ref{inv2}) and (\ref{inv5}),
\beq\label{inv13}
\p_t^2\log \tau (x)=\frac{\tau (x+\eta )\tau (x-\eta )}{\tau^2(x)} \quad \mbox{at $t+\bar t=0$}.
\eeq
This is the turning points condition in terms of the tau-function.
Writing it as
\beq\label{inv13a}
\frac{\ddot \tau(x)}{\tau(x)}-\Bigl (\frac{\dot \tau (x)}{\tau(x)}\Bigr )^2=
\frac{\tau (x+\eta )\tau (x-\eta )}{\tau^2(x)}
\eeq
and comparing the leading singularities of both sides at $x=x_i$, where $x_i$ is any
zero of the tau-function, we obtain the turning points condition (\ref{turn2}).
\square

Comparing (\ref{inv13}) with the 2D Toda equation (\ref{toda}), we can represent it
in the form
\beq\label{inv14}
(\p_{t_1}+\p_{\bar t_1})\p_{t_1}\log \tau (x)=0 \quad
\mbox{or $\,\,\, (\p_{t_1}+\p_{\bar t_1})\xi_1(x)=0$ at $t_1+\bar t_1=0$}
\eeq
which agrees with (\ref{inv17}).

\section{Conclusion}

The main result of this paper is introduction of a new integrable hierarchy which we
have called the constrained Toda hierarchy 
or simply C-Toda hierarchy. It is obtained from the 2D Toda lattice by imposing
a constraint on the two Lax operators of the latter.
The constraint
is invariant with respect to only a ``half'' of the hierarchical time flows, so the other
half of the time variables should be ``frozen'' (fixed to zero values).
The story is to much extent analogous
to the way in which the CKP hierarchy is obtained from the KP hierarchy. The analogy
also manifests itself in the construction of the tau-function of the C-Toda hierarchy.

A related result concerns elliptic solutions to the C-Toda hierarchy and their relation
with the elliptic Ruijsenaars-Schneider model. We have shown that zeros of the tau-function
of the ellipic solutions move as Ruijsenaars-Schneider particles restricted to a
half-dimensional submanifold in the phase space corresponding to {\it turning points}.
In this respect, too, the situation is analogous to the CKP case, where the dynamics
of poles of elliptic solutions is the Calogero-Moser dynamics restricted to the submanifold
of turning points, i.e. points with zero momenta, as is shown in \cite{KZ21}.

\section*{Acknowledgments}

\addcontentsline{toc}{section}{\hspace{6mm}Acknowledgments}

The research has been funded within the framework of the
HSE University Basic Research Program.

\end{document}